# A Highly Compact Direct-Injection Power-Flow Controller and Line-Voltage Regulator with Shared Magnetics and Partial-Power Conversion for Full-Power Control

Davood Keshavarzi, Alexander Koehler, and Stefan M. Goetz

*Abstract*—An increasing integration of photovoltaic units, electric vehicle chargers, heat pumps, and energy storage systems challenges low-voltage power grids and can cause voltage range violation, loss of stability, (local) overload of lines, and power management problems. Research suggested universal power-flow control (UPFC) to solve power management problems. In contrast to bulky, slow, and costly conventional UPFCs with their shunt and series transformers, this paper presents a highly compact and current-dense power-flow controller, which can serve between different feeders in the low-voltage power grids. The enabler is a systematic combination of silicon carbide (SiC) with silicon (Si) transistors and a strict partial-power topology built around a multi-active bridge. The circuit links an active-front-end converter as a shunt stage through a multi-active-bridge converter bidirectionally with low-voltage series-injection modules floating with their respective phases. The topology can use small power to control high currents through the low-voltage series-injection modules. The multi-active bridge serves as a multi-input-output power router that exchanges energy between all elements. We assess the design as well as the implementation considerations of the proposed power-flow controller mathematically and verify its performance in simulation and real systems.

## I. Introduction

THE expansion of low-scale photovoltaic units, heat pumps, energy storage systems, and electric vehicle charging stations with intermittent power injection or extraction behavior causes extreme challenges for reliability and power quality in low-voltage (LV) and last-mile distribution grids [1]. These high-power and typically rapidly controlled devices affect the voltage. The massive solar integration of recent years even inverts the power-flow direction to sometimes multiple times the maximum rated load [2]. The resulting reverse power flow can overload lines and locally increase the grid voltage beyond the standard limit of ±10% deviation, as for instance set by the European Standard EN50160 [3]. Therefore, voltage rises and fluctuations need compensation. Conventional devices such as on-load tap changers [4], step voltage regulators [5], switched shunt capacitors [6], and reactive power management [7, 8] can either lower or raise the voltage across an entire feeder. However, in some places, over- and under-voltage conditions can occur simultaneously. These fluctuations along the feeder cannot be adequately mitigated by these traditional devices, or only if used in large quantities. Houses in recent residential development areas nowadays typically incorporate large photovoltaic installations and may be connected at the end of an existing feeder that previously supplied older buildings with small or no solar power. The voltage violation can therefore not be solved at the feeder. The combination of high-power generation (solar) and loads (vehicle chargers and heat pumps) may further lead to local power flows that are not detectable at the feeder and can locally overload a line.

The flexible alternating current transmission system (FACTS) offers a catalog of power-electronic devices to improve efficiency, power control, and voltage regulation at the high-voltage level [9]. The unified power flow controller (UPFC) is one of those devices and can be adapted for LV grids to condition the power flow and quality by controlling parameters such as voltage amplitude and phase angle simultaneously or selectively [10, 11]. A UPFC consists of two partial-power back-to-back voltage source converters connected to the grid through a shunt and a series transformer, both operating at grid frequency. However, these transformers can be massive. Particularly, the injection transformers have to manage the entire current of a feeder.

An application for power quality and flow control in LV grids is the soft open point (SOP). SOPs are power electronic devices that are usually placed at normally open points in distribution grids to provide power and voltage control [12]. Most SOP circuits implement back-to-back converters, which push power from one ac feeder through a dc link and back to another ac feeder. In contrast to UPFCs, however, back-to-back converters need to handle the full line voltage and the full current with their transistors.

Advances in semiconductor technology, not only wide-bandgap materials but also more advanced processes and structures in silicon, enable direct grid connection of power electronics with high voltage or current ratings. Smart circuit topologies, on the other hand, can reduce size and power loss. Lu et al. previously proposed a direct-injection concept for UPFC and SOP operation (Fig. 1). The circuit involves two stages of shunt and series converters directly connected to the grid, eliminating bulky transformers [13]. A hybrid type is also presented with GaN devices to avoid large discrete high-current filtering inductors [14]. These concepts massively reduce the size compared to the state of the art. However, they still rely on three full-size isolated bidirectional dc-dc converters, one for each floating module, which dominate the overall cost and size.

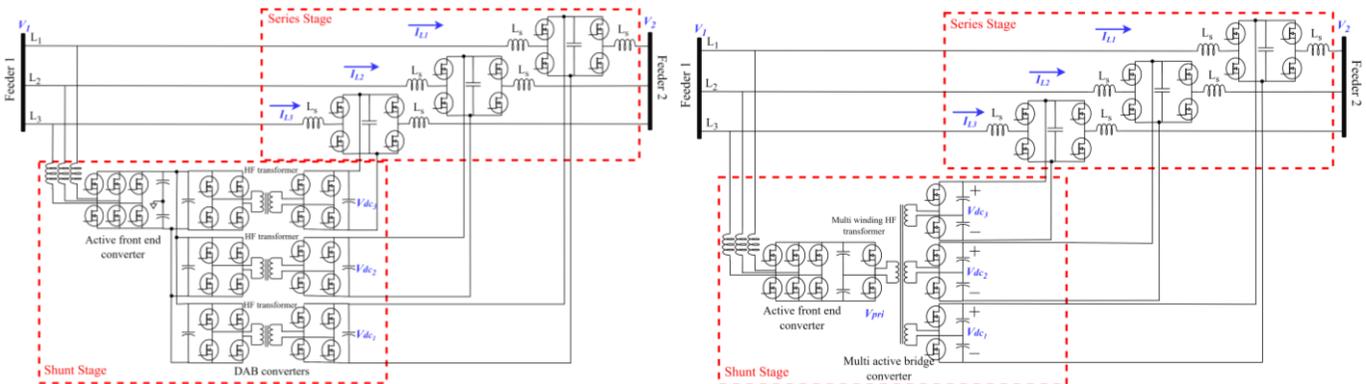

Fig. 1. Circuit diagram of the power-flow controller presented by Lu et al. [13] (left) and with the proposed compact MAB concept (right).

In this paper, we propose a more compact power-flow controller with optimized control. It eliminates low-frequency transformers in favor of directly grid-tied fully electronic shunt and series stages. More importantly even, it does not require independent isolated dc-dc converters for each phase. A multi-active-bridge (MAB) converter forms the centerpiece of the proposed power-flow controller. The MAB converter galvanically isolates the connected elements and freely exchanges power between the injection modules and the active front-end serving as a shunt module. Moreover, we present an optimized controller for each stage for stable operation during unbalanced conditions. This design shares all magnetics and reduces the semiconductor count so that the size and the weight decrease further.

Beyond actuating the voltage, the possibility of controlling power flow in LV grids without remote-throttling distributed power generation or grid reconfiguration can improve the grid performance. In loops or segments supplied through multiple feeders, e.g., as soft open points, the proposed power-flow controller can change the current flows on the line in such a way that thermal limits are not exceeded, stability margins increase, and contractual requirements are fulfilled without curtailment. Although the circuit can use high-level control approaches from SOPs and UPFCs, the main contributions of this paper focus on the topology as follows:

- The proposed circuit employs low-voltage high-current components in the series stage and high-voltage low-current semiconductors in the shunt stage. Whereas back-to-back converters need semiconductor devices with full voltage and current ratings, these components with just high voltage or high current are easily accessible and have a relatively fast switching speed.
- In contrast to back-to-back converters, the proposed circuit needs to convert a fraction of line power. By injecting a small voltage, it can push enormous current. The typical small deviation in voltage and low impedance in LV feeders needs only low voltages—a few volts for LV grids—to control power flow. Since the series-injection modules are floating with the phase voltage, they can only employ low-voltage semiconductors.
- Compared to conventional UPFCs, which typically use series (and shunt) injection transformers, the proposed circuit eliminates bulky low-frequency transformers, which add cost and size to the system. With dynamic high-power demands as in electric vehicle chargers and renewable energy sources such as solar units, low-voltage distribution grids are having trouble controlling power flow. This novel circuit is beneficial in these situations.
- Compared to other universal direct-injection power-flow controllers that rely on isolated dc-dc converters in each phase, the proposed circuit with the MAB concept links the series-injection modules through one shared magnetic galvanic isolation. This concept not only reduces the number of semiconductor components in the system but also avoids core saturation due to the split dc-link in the MAB's terminals.
- Compared to a transformer-based system, the proposed circuit has faster dynamics and higher bandwidth.

The following Section II discusses modeling and a control strategy for stable operation. Section III details active and reactive power management as well as power flow control by the proposed circuit. Section IV presents analysis and simulation results. Corresponding experimental results in Section V confirm the performance of the proposed power-flow controller. Section VI analyzes the power loss of the circuit. Section VII studies the dynamics before Section VIII concludes the paper.

## II. SYSTEM CONFIGURATION

Figure 1 illustrates the concept of the proposed power-flow controller. Overall, it consists of three essential stages: (1) three high-current series-injection modules, (2) a grid-voltage shunt converter, and (3) an MAB link. The series-injection module with two inductors on both sides is repeated for each phase. The series inductors can be configured with small capacitors to form a second-order filter. The series-injection module plays the key role in regulating the injected voltage and/or resulting current in the line with small floating voltage. The second stage is one shared shunt converter, which provides a well-regulated dc-bus voltage. The MAB converter isolates the supply as well as the

· 2 ·

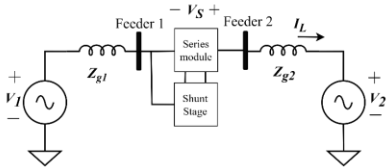

Fig. 2. Single-line circuit diagram of the series-injection converter between two feeders.

three series-injection modules from each other and manages the power exchange between the phases.

For maximum flexibility, all converters are bidirectional to supply the series modules but also extract power. Grid-voltage SiC transistors allow a direct grid connection of the shunt converter to the LV feeder without a bulky transformer at grid frequency. Wide-bandgap SiC transistors further allow a high switching rate for a small grid filter, low switching loss, and a fast response with proper control.

The differential voltage of the series modules can control the active and reactive power flow. Due to typically small voltage deviation and low impedance in meshed or radial LV distribution feeders, only a few volts difference (<15% of the grid voltage) can drive extreme current through two grid sectors. This advantage favors low-voltage Si field-effect transistors with very high current capability and even allows the proposed power-flow controller to participate in fault current supply with thousands of amps by appropriate parallel configuration.

*A. Series stage*

The series-injection modules provide the main function of the power-flow controller by regulating an ac voltage with controllable magnitude and phase angle via an H-bridge module connected in series with the grid lines. The Series modules' dc-link voltages ($V_{dc1}$, $V_{dc2}$, $V_{dc3}$) have the same level since the phase voltages change only slightly in amplitude. The H-bridge module is configured with small series inductors on both sides. Figure 2 represents the single-phase circuit diagram of the series stage placed between two feeders with given line impedances. The series-injection module can inject regulated current from one side to the other in any direction. The line current can be obtained as

$$I_g = \frac{\vec{V_1} + \vec{V_S} - \vec{V_2}}{Z_g}, \quad (1)$$

where $V_1$ and $V_2$ are the feeders' open-circuit voltages, $V_S$ is the series-injection H-bridge module's voltage, and $Z_g$ is the total line impedance ($Z_{g1} + Z_{g2}$). From (1), it is obvious that the injected current can be controlled through the series voltage $V_S$. Since the series voltage $V_S$ is much lower than the feeder's voltage, only a small portion of the power is injected or extracted by the series-injection modules (<15% of the full line's power). As it is assumed that the dc-link voltage is stiff, the series-injection modules, acting as current-controlled inverters, can be modeled as linear time-invariant systems. For one phase, we can derive equations for $I_d$ and $I_q$ in the d–q rotating frame as

$$\begin{aligned} V_{Sd} &= V_{2d} - V_{1d} - \omega_g L_g I_q, \\ V_{Sq} &= V_{2q} - V_{1q} + \omega_g L_g I_d, \end{aligned} \quad (2)$$

where $\omega_g$ and $L_g$, respectively, are the grid frequency and the total line inductance [15]. d–q transformation facilitates the use of a linear PI controller for current regulation. Figure 3 depicts the control block diagram of the series module that can be repeated for each phase. With this control strategy, we can control the current (or power flow) for each phase independently and even during imbalance.

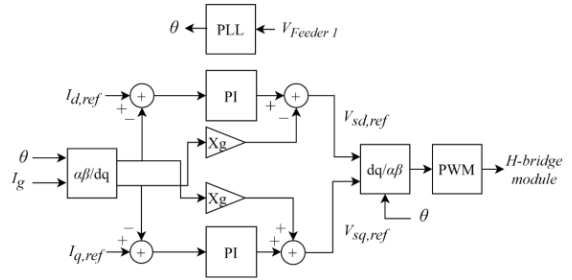

Fig. 3. Control block diagram of the series-injection modules.

*B. Shunt stage*

As discussed in the previous section, the series stage needs a stiff floating dc voltage supply for each H-bridge module for accurate power-flow control. We proposed an active-front-end (AFE) converter in combination with an MAB converter. The AFE converter works as an interconnecting bridge between the ac mains and the dc side with various advantages, such as bidirectional power flow, unity power factor or targeted reactive current injection, and low total harmonic distortion [16].

The MAB converter as an expansion of a dual-active-bridge (DAB) topology offers high power capability. The DAB family is known for its high efficiency levels, ability to handle a broad range of input and output voltages, effective isolation, compactness, and low weight [17, 18]. However, three separate DAB converters, one per phase, can become bulky and costly. Therefore, we decided to merge the converters. The proposed configuration features galvanic isolation and the highest efficiency in size and cost of the magnetic core. Also, it employs half bridges and split dc-link capacitors to avoid unwanted core saturation, which can occur, particularly for high-power converters with highly efficient low-loss transformers.

 1) *AFE converter control*

The AFE serves as the power supply and feeds a shared dc bus, which in turn supplies the MAB converter. As the series-injection modules may extract power from the grid, e.g., if they reduce the current that would naturally flow against the direction of the voltage gradient (i.e., from higher to lower voltage), all elements including the AFE converter should be bidirectional to transport the energy accumulating in the modules back to the AFE and feed it back into the grid. It should further be able to manage as well as compensate for imbalance and transients [19].

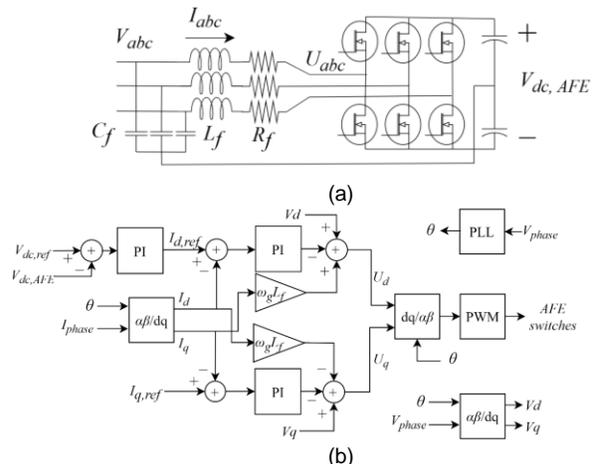

Fig. 4. AFE converter: (a) circuit diagram, (b) control block diagram.

Figure 4(a) presents the circuit diagram of the AFE converter. The combination of a three-leg SiC transistor bridge with an LC filter at the ac side and split dc-bus capacitors presents a capacitor-countered three-phase four-wire inverter. This configuration has two advantages: first, this inverter behaves as three parallel half-bridge converters with decoupled phases and blocked common mode (without neutral inductor). Second, a split dc-bus voltage is provided for the MAB converter without the need for extra high-voltage capacitors. Moreover, current and power control is feasible for each phase by controlling the half-bridges independently with a shared dc bus [20].

We designed a voltage-oriented control with a PI current regulator to control the AFE converter [21-24]. Although other controllers may show a good dynamic response, they often have relatively large steady-state error [25]. The controller transforms the AFE currents into the d–q rotating frame synchronous to the filter capacitor voltage. The advantage of such a control is robustness against grid impedance variation. The phase voltage can be described in the d–q rotating frame by

$$V_{d,C_f} = U_d - \omega_g L_f I_q + K_P\left(1 + \frac{1}{\tau_i s}\right)(I_d^{\text{ref}} - I_d),$$
$$V_{q,C_f} = U_q + \omega_g L_f I_d + K_P\left(1 + \frac{1}{\tau_i s}\right)(I_q^{\text{ref}} - I_q).$$
(3)

Figure 4(b) presents the control block diagram of the AFE converter. The control concept consists of two control loops; the outer loop regulates the dc-bus voltage shared by all three phases, while the inner loop regulates the ac-side phase current. The dc voltage of the AFE converter should be high enough (i.e., $\geq 700$ V) to reduce the current distortion [26, 27]. The proportional gain ($K_P$) and the integral constant ($\tau_i$) can be calculated as

$$K_P = \frac{L_f}{1.5\, a\, T_s},$$
$$a = \frac{1}{\frac{\pi}{2} - \varphi_m},$$
$$\tau_i = \frac{L_f}{R_f},$$
(4)

$T_s$ : sampling time,

where $\varphi_m$ is the phase margin of the controller [28]. In control theory, a phase margin greater than 45 degrees is usually considered to have stable operation. Therefore, $a > 2$ is chosen as stability criteria in practical application.

2) *MAB converter control*

The MAB converter is responsible for regulating the dc-link voltage of each H-bridge module and galvanically isolates the modules from the remaining circuit and ground so that they can float with their respective phase of a power line. It comprises a high-voltage half-bridge inverter at the primary and three low-voltage half-bridge inverters at the secondary side coupled to a single multi-winding high-frequency transformer as shown in Fig. 5(a).

The voltage regulation relies on cross-coupling of the matrix power flow in the multi-winding transformer. Instead of time-sharing control, which would reduce the efficiency under heavy-load conditions [29], we use phase-shift control to determine the power sharing between outputs [30]. The power flow in the multi-winding transformer reflects the injected or extracted power to and from the lines. Furthermore, it has similarities to a meshed inductively dominated grid, where power flows into the individual lines is likely determined by the phase shift. Fig. 5(b) illustrates the phase-shifted modulation, Fig. 5(c) the star and delta configuration of the high-frequency transformer. A modification in the phase shift of one bridge affects the power transfer of all other bridges. The delta configuration can simplify the analysis of the current flows through the inductors. The power average of the $i$-th bridge over one switching period follows

$$P_i = \sum_{i \neq j} \frac{V_i n_{ij} V_j}{8\pi^2 f_{\text{sw}} L_{ij}}(\varphi_i - \varphi_j)(\pi - |\varphi_i - \varphi_j|),$$
(5)

where $V_i$ and $V_j$ denote the dc bridge voltages, $n_{ij}$ is the turn ratio between two bridges, $L_{ij}$ represents the equivalent inductance from two bridges, $f_{\text{sw}}$ refers to the switching frequency, and $\varphi_i$ as well as $\varphi_j$ are the phase-shift angles of two bridges [31]. The equivalent inductance between bridge $i$ and bridge $j$ in the Delta model can be calculated as a function of the inductance in the star model as

$$L_{ij} = L_i L_j \left(\frac{1}{L_m} + \sum \frac{1}{L_k}\right).$$
(6)

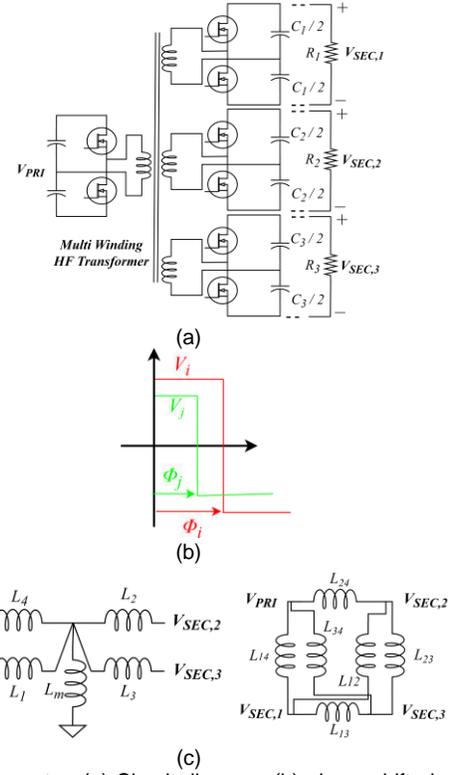

Fig. 5. MAB converter: (a) Circuit diagram, (b) phase-shifted modulation, (c) star and delta model of the HF transformer.

Usually, the primary voltage ($V_{\text{PRI}}$) is set as the reference, and its phase is zero. One of the challenges of MAB is that individual control of one bridge will unbalance all other bridges. We can calculate the corresponding phase-shift angles [$\Phi_1\ \Phi_2\ \Phi_3$] of the secondary bridges for a feasible power-flow matrix numerically (specifically Newton–Raphson). The average output current of each half bridge follows

$$I_i = \frac{1}{8\pi^2 f_s} \sum_{i \neq j} \frac{n_{ij} V_j}{L_{ij}} (\varphi_i - \varphi_j)(\pi - |\varphi_i - \varphi_j|). \quad (7)$$

Linearization of (7) around the operating point yields a small-signal transfer function of the current as

$$\hat{i}_i = \frac{1}{8\pi^2 f_s} \times \left[ \hat{\varphi}_i \sum_{i \neq j} \frac{n_{ij} V_j}{L_{ij}} (\pi - 2|\Phi_i - \Phi_j|) \right.$$

$$\left. - \sum_{i \neq j} \hat{\varphi}_j \frac{n_{ij} V_j}{L_{ij}} (\pi - 2|\Phi_i - \Phi_j|) \right] \quad (8)$$

$$= K_{\varphi i} \hat{\varphi}_i - \sum_{i \neq j} K_{\varphi ij} \hat{\varphi}_j.$$

By considering the impedance of the dc link, the small signal voltage transfer function follows as

$$\hat{v}_i = Z_i \times \hat{i}_i = \frac{R_i}{R_i C_i s + 1} \left( K_{\varphi i} \hat{\varphi}_i - \sum_{i \neq j} K_{\varphi ij} \hat{\varphi}_j \right) \quad (9)$$

$$= G_{vi} \hat{\varphi}_i - \sum_{i \neq j} G_{vij} \hat{\varphi}_j,$$

where $C_i$ and $R_i$ are, respectively, the dc link capacitor and equivalent load of the $i$-th secondary bridge.

The controller must achieve good tracking of the reference point with no steady-state error, as well as a fast transient response. In consequence, the controller should have high gain at dc and a high cross-over frequency. Since the reduced model is of first order inherently and the controller regulates a dc quantity, a PI controller is sufficient for high performance in output voltage regulation. On the other hand, the ratio of $G_{vij} \forall (i \neq j)$ to $G_{vi}$ represents the degree of cross-coupling between individual control loops. The following decoupling term should be added to the output of the phase controller of each secondary:

$$\Delta \varphi_i = \sum_{i \neq j} \frac{K_{\varphi ij}}{K_{\varphi i}} \varphi_j. \quad (10)$$

Figure 6 depicts the control block diagram of the MAB converter for secondary #1 with the decoupling term. This control diagram can be applied to other secondaries.

The open-loop transfer function of the MAB's controller is defined as

$$G_{OL} = K_P \left( 1 + \frac{1}{\tau_i s} \right) e^{-sT_d} K_{\varphi i} \frac{R}{RCs + 1}, \quad (11)$$

where $K_P$, $\tau_i$, and $T_d$ are, respectively, the proportional gain, integral time-constant, and digital delay. The controller bandwidth depends on the cross-over frequency and the desired phase margin ($\varphi_m$). It can be calculated as

$$\angle G_{OL}(j\omega_c) = -\pi + \varphi_m \Rightarrow \omega_c = \frac{\frac{\pi}{2} - \varphi_m}{T_d}, \quad (12)$$

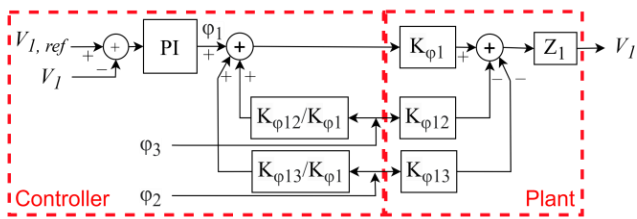

Fig. 6. Control block diagram of MAB for secondary #1.

### TABLE I
### MAB PARAMETER VALUES.

| Symbol | Description | Value |
| --- | --- | --- |
| $V_{pri}$ | Primary voltage | 800 V |
| $V_{sec}$ | Secondary voltage | 50 V |
| $f_{sw}$ | Switching frequency | 100 kHz |
| $f_s$ | Sampling frequency | 10 kHz |
| $L_s$ | Inductance between primary and secondary | 15 µH |
| $C$ | Secondary dc bus capacitor | 200 µF |
| $\omega_c$ | Controller bandwidth | 2 kHz |
| $K_P$ | Proportional gain | 0.040 |
| $\tau_i$ | Integral time constant | 1 ms |

where $RC\omega_c \gg 1.0$ and $\tau_i \omega_c = 10$. The proportional gain $K_P$ follows from setting the magnitude of the open-loop transfer function to unity at the cross-over frequency as

$$|G_{OL}(j\omega_c)| = 1 \Rightarrow K_P = \frac{C\omega_c}{K_{\varphi i}}. \quad (13)$$

The proportional gain depends on the $K_{\varphi i}$ term, which varies significantly with the phase-shift operation point. Therefore, the optimized proportional gain calculated for the nominal point will not present the same performance over the entire operating range. To solve this problem, sensors are used at the MAB's terminals to measure output powers. Then, the MAB's controller can determine the equivalent load ($R_i$) to calculate feasible $K_{\varphi i}$ at the operating point. Table I lists the MAB converter parameters.

## III. ACTIVE AND REACTIVE POWER MANAGEMENT WITH POWER FLOW CONTROL

The series-injection modules' output voltage determines the power-flow control for the proposed concept. We study the performance of the proposed power-flow controller with P–Q load and two active feeders. As other power electronics devices, the proposed power-flow controller has an operating area that guarantees safe operation. The series module's voltage can be written as

$$V_S = r\vec{V}_1 e^{j\gamma}, \quad (14)$$

where $0 < r < r_{max}$ and $0 < \gamma < 2\pi$. Ignoring over-modulation, $r_{max}$ is equal to $\frac{V_{dc}}{\sqrt{2} V_1}$.

### A. Feeding a P–Q load

The proposed power-flow controller can provide numerous functions. It can, for instance, feed a load or a downstream branch with several loads and grid-following sources. In such a scenario, it can inject and such completely compensate for the reactive power of a load on one side of the power-flow controller. This mode entirely blocks reactive power from passing through so that the line upstream does not experience any phase shift caused by downstream lines, loads, or sources. Figure 7 illustrates the equivalent single-phase circuit feeding a P–Q load. The active and reactive power shares of the feeder can be written as

$$S_1 = V_1 I_L^* = \quad (15)$$

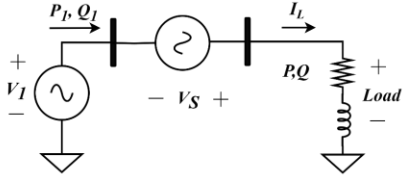

Fig. 7. Single-line circuit diagram of series module feeding a P–Q load.

$$\begin{cases} P_1 = \dfrac{V_1^2}{Z_L}\cos\varphi + \dfrac{rV_1^2}{Z_L}\cos(\varphi-\gamma) \\ Q_1 = \dfrac{V_1^2}{Z_L}\sin\varphi + \dfrac{rV_1^2}{Z_L}\sin(\varphi-\gamma). \end{cases}$$

Fully reactive and active load compensation by the series-injection module leads to

$$\begin{aligned} \sin\varphi &= r\sin(\varphi-\gamma) \\ \sin\left(\varphi-\dfrac{\pi}{2}\right) &= r\cos(\varphi-\gamma). \end{aligned} \quad (16)$$

Supposing the maximum amplitude without over-modulation for the series modules, the operating area for the P–Q load can be obtained as

$$\begin{aligned} -\dfrac{V_{dc}}{\sqrt{2}V_1} &\le \sin\varphi \le \dfrac{V_{dc}}{\sqrt{2}V_1} \Rightarrow \\ -\sin^{-1}\left(\dfrac{V_{dc}}{\sqrt{2}V_1}\right) &\le \tan^{-1}\left(\dfrac{Q}{P}\right) \le \sin^{-1}\left(\dfrac{V_{dc}}{\sqrt{2}V_1}\right) \\ -\dfrac{V_{dc}}{\sqrt{2}V_1} &\le \sin\left(\varphi-\dfrac{\pi}{2}\right) \le \dfrac{V_{dc}}{\sqrt{2}V_1} \Rightarrow \\ -\sin^{-1}\left(\dfrac{V_{dc}}{\sqrt{2}V_1}\right) &\le \dfrac{\pi}{2} - \tan^{-1}\left(\dfrac{Q}{P}\right) \le \sin^{-1}\left(\dfrac{V_{dc}}{\sqrt{2}V_1}\right). \end{aligned} \quad (17)$$

### B. Connection between two feeders

The proposed power-flow controller can further be placed between two feeders with different amplitudes and phases as shown in Fig. 2. In this condition, the modules control the injected power to Feeder 2. By considering the equivalent circuit diagram in Fig. 8, the injected power to Feeder 2 can be obtained as

$$S_{inj} = V_2 I_g^* = \qquad (18)$$

TABLE II
SYSTEM PARAMETER VALUES.

| Symbol | Description | Value |
|---|---|---|
| $V_g$ | Grid voltage | 400 V rms |
| $f_g$ | Grid frequency | 50 Hz |
| $Z_g$ | Phase impedance | 20 + j50 mΩ |
| $V_{dc,AFE}$ | AFE dc voltage | 800 V |
| $V_{dc}$ | Series-injection converter dc voltage | 50 V |
| $f_{sw}$ | Switching frequency | 100 kHz |
| $L_s$ | Series inductance | 100 µH |
| P–Q load | Case 1 | 40 kW+j5.621 kVar |
| Feeder 2 | Case 2 | 380 V rms, ∠0° |
|  | Case 3 | 400 V rms, ∠8° |
| Unbalance | Case 4 | (u, v, w) = (200, 230 250) V rms 30 kW + j4.5 kVar |

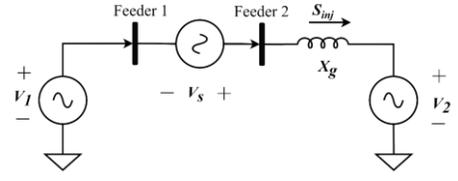

Fig. 8. Single-line circuit diagram of series module connected between two feeders.

$$\begin{cases} P_{inj} = \dfrac{V_1 V_2}{X_g}\sin(\theta_1-\theta_2) + \dfrac{rV_1 V_2}{X_g}\sin(\theta_1-\theta_2+\gamma) \\ Q_{inj} = \dfrac{V_1 V_2}{X_g}\cos(\theta_1-\theta_2) + \dfrac{rV_1 V_2}{X_g}\cos(\theta_1-\theta_2+\gamma) \\ \qquad -\dfrac{V_2^2}{X_g}. \end{cases}$$

As in the previous section, we can have full compensation for active and reactive power. Therefore, an operation area can be obtained for dedicated active and reactive power as

$$\begin{aligned} -\dfrac{V_{dc}}{\sqrt{2}V_1} &\le \sin(\theta_1-\theta_2) \le \dfrac{V_{dc}}{\sqrt{2}V_1} \Rightarrow \\ -\sin^{-1}\left(\dfrac{V_{dc}}{\sqrt{2}V_1}\right) &\le \theta_1-\theta_2 \le \sin^{-1}\left(\dfrac{V_{dc}}{\sqrt{2}V_1}\right). \\ -\dfrac{V_{dc}}{\sqrt{2}} &\le V_1\cos(\theta_1-\theta_2) - V_2 \le \dfrac{V_{dc}}{\sqrt{2}} \Rightarrow \\ -\dfrac{V_{dc}}{\sqrt{2}} &\le V_1 - V_2 \le \dfrac{V_{dc}}{\sqrt{2}}. \end{aligned} \quad (19)$$

Equation (19) sets the constraints for maximum amplitude and phase differences, which the power-flow controller can bridge for a certain module's voltage level. Beyond those limits, the system would bypass.

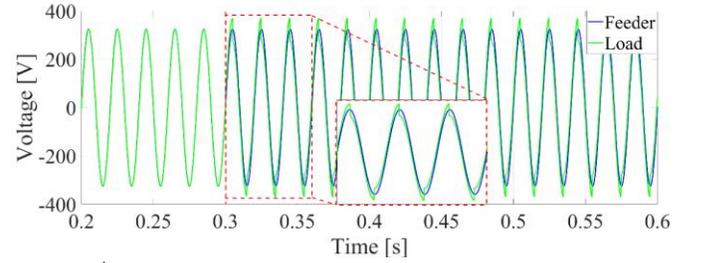
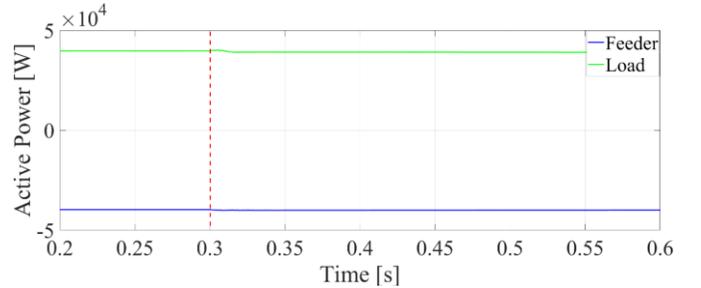
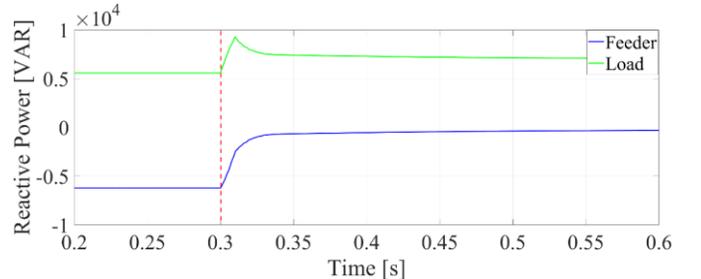

Fig. 9. Voltage, active, and reactive power of P–Q load fed by the power-flow controller.

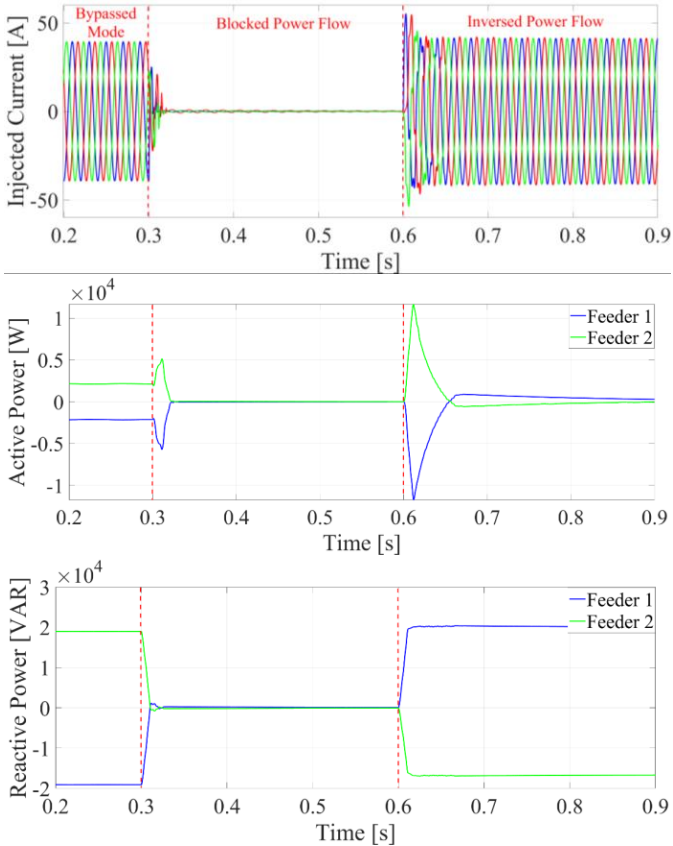
Fig. 10. Line current, active, and reactive power to Feeder 2 in Case 2.

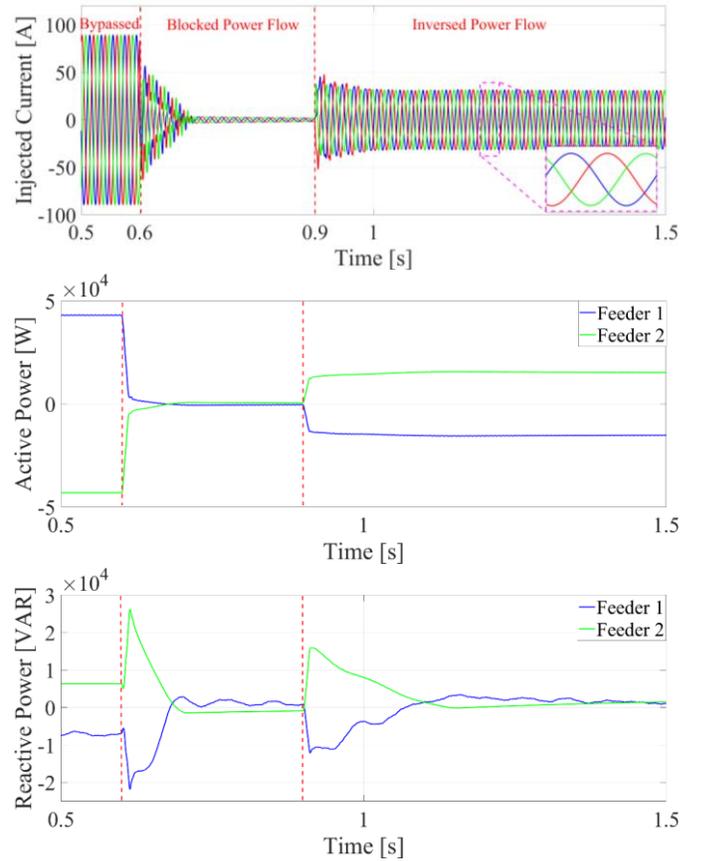
Fig. 11. Line current, active, and reactive power to Feeder 2 in Case 3.

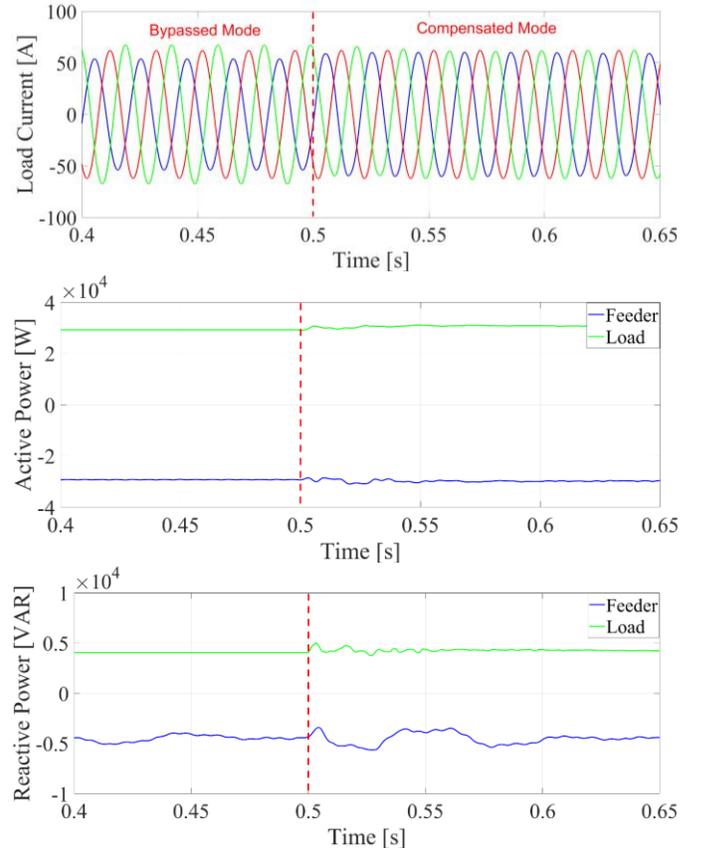
Fig. 12. Line current, active, and reactive power to Feeder 2 in Case 4.

## IV. SIMULATION

We studied the performance of the proposed power-flow controller in MATLAB/Simulink for three scenarios. Table II lists the corresponding system parameters.

### A. P–Q load compensation

In this scenario, we placed the circuit between the grid feeder and a P–Q load with nominal values. According to (17) and the system parameters, the maximum/minimum load angle can be ±8.84° for fully reactive power compensation. Figure 9 graphs the voltage, current, and active/reactive power by bypassing the series-injection modules for Case 1. At $t = 0.3$ s, the series-injection modules turn into active operation. After that time, the proposed power-flow controller compensates for all reactive power.

### B. Two feeders or nodes

One of the primary functions of a power-flow controller is the control of active/reactive power between and/or adapting voltage as well as phase differences of two grid feeders or nodes. The first scenario, Case 2, places the circuit between two feeders with different voltage amplitudes but the same phase. Normally, the current would flow from the higher to the lower voltage. This would also be the default if the modules are in the bypass mode. Figure 10 graphs the power flow and current of Feeder 2. Since the feeders' voltages are different in amplitude and not in phase angle, most of the power is reactive. At $t = 0.3$ s, the series-injection modules turn on and regulate the current down. The series-injection modules cannot only regulate

· 7 ·

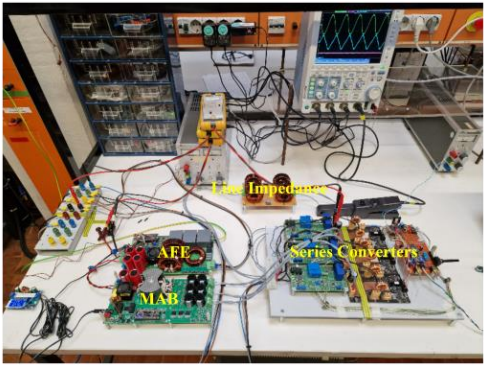

Fig. 13. Experiment setup.

TABLE III
SPECS OF THE EXPERIMENTAL DEMONSTRATOR.

| Symbol | Description | Value |
|---|---|---|
| $V_g$ | grid voltage | 400 V rms |
| $f_g$ | grid frequency | 50 Hz |
| $R_{Line}$ | line resistance | 40 mΩ |
| $L_{Line}$ | line inductance | 700 µH |
| $V_{dc,AFE}$ | AFE output voltage | 750 V |
| $V_{dc}$ | Series-injection module dc voltage | 50 V |
| $f_{sw}$ | switching frequency | 50 kHz |
| $L_s$ | series inductance | 100 µH |
| AIMBG120R040 | HV switches | 1200 V/70 A |
| IPT015N10N05 | LV switches | 100 V/360 A |

the current down but also push current from the feeder with the lower voltage to the one with the higher voltage (see $t > 0.6$ s).

The second scenario, Case 3, defines two feeders with different phase angles but the same amplitude. The active power flows from Feeder 2 to Feeder 1 when the series-injection modules are bypassed. Figure 11 illustrates the injected power and current to Feeder 2 when the series-injection modules turn on at $t = 0.6$ s. It is obvious that the proposed power-flow controller can regulate both active and reactive power. As demonstrated before, the power-flow controller is capable of driving active power to Feeder 2 against the natural power-flow direction.

### C. Unbalanced network

Unbalanced conditions are common in LV grids. Therefore, we studied Case 4, which operates the proposed power in an unbalanced feeder (Fig. 12). We place the circuit between imbalanced Feeder 1 and a P–Q load. At $t = 0.5$ s, the circuit turns from bypass to compensation and balances the current in the three phases for the load feeder.

## V. EXPERIMENTS

We implemented a physical prototype to study the performance of the power-flow controller experimentally. Table III summarizes the specs of the setup. We used a grid simulator (IT7915-350-90) and inductors to simulate an LV system. We assumed a meshed power grid in which the proposed circuit is placed between two feeders that connected to the medium voltage grid through a transformer. Silicon-transistor power modules with high current capability (Infineon IPT015N10) formed the series stage for the three phases. The testbench configuration is shown in Fig. 13. Also, Fig. 14 depicts the size comparison between the DAB and the MAB converter as well as the entire experimental setup with three DAB stages (corresponding to the setup with MAB in Fig. 13). The MAB drastically decreases the size of the power-flow controller. This visual impression is backed by the numbers in Table IV, where the number of components, switch rms current, weight, and volume are listed for a power of 15 kW. Both the weight and size of the MAB account for less than one-third of the DABs for the same power.

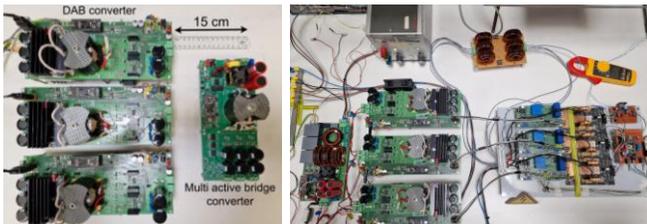

Fig. 14. Size comparison: three DAB converters versus the one MAB converter (left panel), experimental setup with DAB converters corresponding to Figure 14 (right panel).

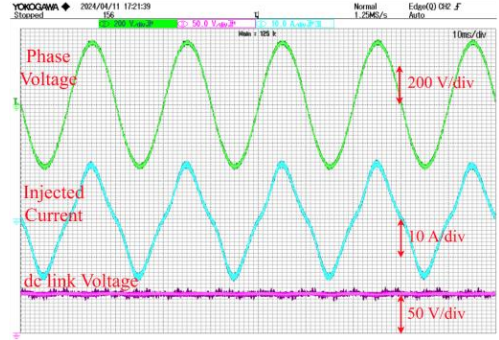

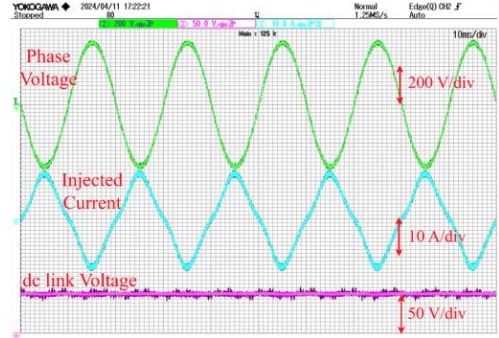

Fig. 15. Experimental results for active power regulation. (a) Forward direction, (b) reverse direction.

We performed experiments to demonstrate the power-flow control function to regulate high power at full grid voltage with the injection of only small voltages and very little converted power through a multi-active bridge with shared magnetics. Figure 15 graphs the grid voltage, the injected current, and the series-injection module capacitor voltage for one phase. The proposed circuit regulates the injected current to ensure that reactive power is kept to zero in this specific test. As seen in Fig. 15(a), the injected current is in phase with the voltage and demonstrates the controller's performance in power regulation. Figure 15(b) in turn regulates the active power in the reverse direction.

We repeated the previous test for reactive power regulation when the active power is kept at zero. Figure 16(a) illustrates reactive power flow with a π/2 phase difference between injected current and voltage. Subfigure (b) reverses the reactive power flow.

We also studied the functionality of the proposed circuit in voltage regulation of a load feeder as the simulation results. We supposed that the circuit feeds a P–Q load with an unbalanced voltage, whereas the series-injection modules are bypassed. The feeder has an rms voltage of 250–230–210 volts. Fig. 17 shows the phase voltages and line current. At $t = t_0$, the circuit is operated and compensates for the load voltage.

TABLE IV
COMPARISON OF MAB AND DAB CONVERTERS FOR A POWER OF 15 KW

|  | MAB converter | 3×DAB converter |
|---|---|---|
| Semiconductor | 2×HV MOSFET<br>6×LV MOSFET | 12×HV MOSFET<br>12×LV MOSFET |
| Capacitor | 2×HV capacitor<br>6×LV capacitor | 3×HV capacitor<br>3×LV capacitor |
| Switch rms current | HV: 33 A<br>LV: 222 A | HV: 7 A<br>LV: 111 A |
| HF transformer | 1 | 3 |
| Weight | 960 g | 3×1250 g |
| Volume | 1932 cm³ | 3×2226 cm³ |

## VI. POWER LOSS ANALYSIS

Power loss analysis of the proposed circuit is more complex since the circuit has more stages and conditions than a normal converter. For a case of injecting power of 15 kW into the grid, the rms current of the lines is 140 A, and the voltage of the series-injection modules is about 35 V. We employed Infineon IPT015N10N05 Si MOSFET for the series-injection modules. The power loss of a semiconductor switch contributed by the conduction and switching is obtained as

$$P_{\text{loss,switch}} = P_{\text{con}} + P_{\text{sw}}$$
$$= R_{\text{on}}I_{\text{rms}}^2 + 0.5V_{\text{dc}}I_{\text{avg}}(T_{\text{on}} + T_{\text{off}})f_{\text{sw}} \quad (18)$$
$$+ 0.5C_{\text{oss}}V_{\text{dc}}^2 f_{\text{sw}},$$

where $C_{\text{oss}}$ is the output capacitance, $T_{\text{on}}$ and $T_{\text{off}}$ respectively are the turn-on and turn-off times of the switch. According to this formula, the series stage's power loss is approximately 293.4 W. Infineon AIMBG120R040M1 and IPT015N10N05 serve in the primary and secondary of the MAB, respectively. The total loss of MAB is determined to be 558.48 W after accounting for the transformer loss of 36.5 W. The same procedure is used for loss of AFE converter as 183.840 W. We should consider the filter loss of 145.2 W, too. Consequently, the system's total loss is around 1180.9 W.

In conventional UPFCs with a series (or shunt) transformer, as presented by Haque et al. [10], the loss of inverters, filters, and transformers contributes to efficiency. There is a 269 W no-load loss in a 15 kVA transformer [32], whereas inverters and filters can have a loss of 515.2 W. Hence, it is possible to compute an overall loss of 784.2 W for conventional UPFCs.

In contrast to earlier art with injection transformers, the loss of the proposed circuit is comparable. It is also extremely compact, which enables it to take advantage of the future rapid advancements in power semiconductors, whereas transformers no longer undergo development.

## VII. DYNAMICS OF THE CIRCUIT

Dynamic characteristics of a control system can be determined by response time and fidelity. In contrast to the proposed power-flow controller and back-to-back converters, conventional transformer-based concepts avoid active filtering due to their limited bandwidth. Direct injection without a series transformer gives access to full converter bandwidth (the bandwidth of the series filters is high enough to compensate for the harmonics and transients). The conventional transformers are designed to support full line current with high saturation, eddy current, and hysteresis loss. Based on the Bertotti model of the transformer [33], the power loss density follows as

$$P_{\text{tra}} = P_{\text{hys}} + P_{\text{eddy}} = \eta B_m^2 f + \frac{\pi^2 t^2 B_m^2}{6\rho} f^2, \quad (19)$$

where $\eta$ is the Steinmetz constant (roughly 15 A m/(V s) here), $B_m$ is flux density (1.5 T), $f$ the current frequency, $t$ is the sheet thickness (27 μm), and $\rho$ is the lamination resistivity (0.48 μΩm) [34]. The transfer function of the transformer can be

$$G_{\text{tra}} = \left|\frac{P_{\text{in}} - P_{\text{tr}}}{P_{\text{in}}}\right|, \quad (20)$$

where $P_{\text{in}}$ is the power to be injected into the grid. From the above equation, the 3-dB bandwidth of the transformer-based power-flow controller is 1 kHz, whereas the proposed circuit presents a higher level, which is defined by the resonance frequency of the series filter. The resonance frequency should be greater than one-third of the sampling frequency ($f_{\text{sampling}}$) for stable operation [35].

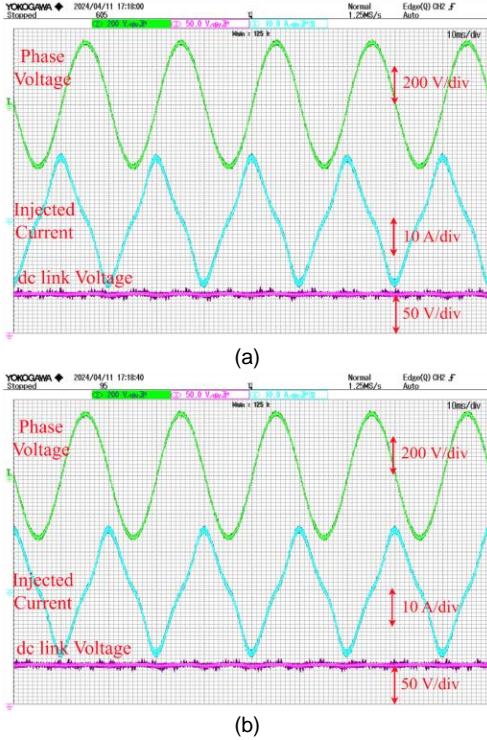

Fig. 16. Experimental results for reactive power regulation. (a) Forward direction, (b) reverse direction.

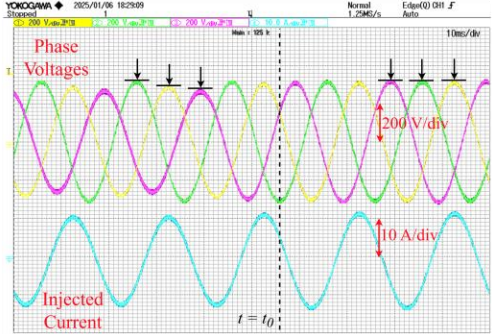

Fig. 17. Experimental result for voltage regulation of unbalanced feeder.

## VIII. Conclusion

Low-voltage grids are facing problems of stability, voltage regulation, and reverse power flow due to an increasing penetration of high-power rooftop solar systems, heat pumps, electric vehicle chargers, and non-grid-supportive energy storage systems. High-voltage grids introduced FACTS devices to solve many issues, but these devices cannot fix the problem of low-voltage grids or are simply technologically not appropriate [36, 37]. Previous suggestions for voltage and power-flow control require bulky and dynamically limited 50 Hz transformers for series-voltage injection. In this paper, we presented a fully electronic modular power-flow controller with direct grid connection without low-frequency shunt/series transformers and a very compact power exchange between modules through shared high-frequency magnetics. The various modules share the same magnetics and reduce size by one-third compared to previous suggestions. We demonstrated the performance and power regulation of the proposed power-flow controller in simulation and experiments.